\documentclass{ckm}                 

\usepackage{graphicx}
\usepackage{isolatin1}

\confname{Workshop on the CKM Unitarity Triangle, IPPP Durham, April
  2003}

\def\eq#1{Eq.\ (\ref{#1})}
\def\eqs#1#2{Eqs.\ (\ref{#1}) and (\ref{#2})}

\def\tab#1{Table \ref{#1}}

\def\sec#1{Section \ref{#1}}

\def\fig#1{Figure \ref{#1}}

\def\l{\left}
\def\r{\right}

\newcommand{\be}{\begin{equation}}
\newcommand{\ee}{\end{equation}}
\newcommand{\bea}{\begin{eqnarray}}
\newcommand{\eea}{\end{eqnarray}}

\newcommand{\nn}{\nonumber}
\newcommand{\ra}{\rightarrow}

\newcommand{\cO}{\mathcal{O}}

\def\ord#1{\cO\l(#1\r)}

\newcommand{\la}{\langle}
\renewcommand{\ra}{\rangle}

\newcommand{\gev}{\mathrm{GeV}}

\newcommand{\fbinv}{\mathrm{fb}^{-1}}
\newcommand{\ps}{\mathrm{ps}}

\def\lsim{\
\lower-1.2pt\vbox{\hbox{\rlap{$<$}\lower5pt\vbox{\hbox{$\sim$}}}}\ }
\def\gsim{\
\lower-1.2pt\vbox{\hbox{\rlap{$>$}\lower5pt\vbox{\hbox{$\sim$}}}}\ }

\title{CKM matrix elements from tree-level decays and 
$b$-hadron lifetimes\thanks{Based on Working Group I's summary
talk at the 2nd Workshop on the CKM Unitarity Triangle, 
IPPP Durham, 5-9 April, 2003.}}

\author{Laurent Lellouch\thanks{Supported in part by
EU HPP contracts HPRN-CT-2000-00145 and HPRN-CT-2002-00311.}\ \ \ {\em for
the convenors of Working Group I 
(Elisabetta Barberio (SMU), Lawrence Gibbons
\newline\phantom{X}\hspace{10cm}(Cornell), Toru Iijima (Nagoya), L.L.) }}

\address{Centre de Physique Théorique, Case 907, 
CNRS Luminy, F-13288 Marseille Cedex 9, France}

\begin{document}

\begin{abstract}
We give an updated summary of the topics covered in Working Group I of
the 2nd Workshop on the CKM Unitarity Triangle, with emphasis on the
results obtained since the 1st CKM Workshop
\cite{Battaglia:2003in}. The topics covered include the measurement of
$|V_{ub}|$, of $|V_{cb}|$ and of non-perturbative Heavy Quark
Expansion parameters, and the determination of $b$-hadron lifetimes
and lifetime differences.
\end{abstract}

\maketitle


\section{Introduction}
\label{sec:introduction}

Uncovering the origin of flavor mixing and CP violation is 
one of the main goals in elementary particle physics today. As
part of this program, the constraining of the CKM unitarity triangle
through the redundant measurement of its angles and sides plays a
central rôle. The aim of the present summary is to review the state
of the art regarding the measurement of the sides $|V_{ub}|$ and
$|V_{cb}|$, as it was discussed in Durham at the 2nd Workshop on the
CKM Unitarity Triangle, taking into account developments which
have occurred since then.

The determination of $|V_{ub}|$ and $|V_{cb}|$ from inclusive decays
relies heavily on the Heavy Quark Expansion (HQE). Many of the
assumptions underlying these calculations can be tested by comparing
HQE predictions for $b$-hadron lifetimes to their experimental
values. Such comparisons are also useful for testing lattice
calculations of non-perturbative hadronic matrix elements which also
enter these predictions. The study of these lifetimes has thus
been included in the subjects covered by our working group.  Of
course, lifetimes are important in themselves. They are required to
convert branching fractions into the rates necessary for the
determination of $|V_{ub}|$ and $|V_{cb}|$. Moreover, lifetime
differences of neutral $B_{s}$ and $B_{d}$ mesons may be helpful for
uncovering new physics.

Spectral moments of inclusive $B$-decay spectra are also covered,
as they too play an important rôle in testing the HQE. They further provide
experimental determinations of many of the non-perturbative parameters
which appear in these expansions. As the latest spectral moment studies
described below show, we are entering a new era, where the
currently achieved experimental precisions require a much tighter
interplay between experiment and theory. A similar interaction is
taking place in the exclusive determination of $|V_{ub}|$, as made
clear by the latest CLEO results \cite{Athar:2003yg}, where
non-perturbative methods such as light-cone sum rules and lattice QCD
are being put to serious test.

The remainder of this summary is organized as follows. In
\sec{sec:inclusiveVub} we review inclusive determinations of
$|V_{ub}|$. Moments of inclusive $B$-decay spectra are discussed in
\sec{sec:moments}, while the measurement of $|V_{ub}|$ from exclusive
semileptonic $B$ decays is the subject of
\sec{sec:exclusiveVub}. In \sec{sec:exclusiveVcb} we review exclusive 
determinations of $|V_{cb}|$
and $b$-hadron lifetimes and lifetime differences in \sec{sec:lifetimes}. We end
with a brief conclusion in \sec{sec:ccl}.

\section{Inclusive determinations of $|V_{ub}|$}
\label{sec:inclusiveVub}

\subsection{Inclusive $|V_{ub}|$: theory}

The status of theoretical calculations relevant for the determination
of $|V_{ub}|$ from inclusive $B\to X_u\ell\nu$ decays is nicely
summarized in the contribution by Luke \cite{Luke:2003nu}. The main
problem is the need for severe phase-space experimental cuts to
eliminate the approximately 100 times larger $B\to X_c\ell\nu$
background: these cuts tend to destroy the convergence of the HQE used
to describe these decays. Theoretical expressions for a variety of
cuts exist:
\begin{itemize}
\item
$E_\ell>\frac{m_B^2-m_D^2}{2m_B}$, where $E_\ell$ is the lepton energy;
\item
$m_X<m_D$, where $m_X$ is the invariant mass of the final hadronic system;
\item
$q^2>(m_B-m_D)^2$, where $q$ is the four-momentum of the leptons;
\item
combined $(q^2,\,m_X)$ cuts.
\end{itemize}
As detailed in \cite{Luke:2003nu}, all of these cuts have advantages and disadvantages, some
experimental, others theoretical. Because the different methods for measuring
$|V_{ub}|$ have different sources of uncertainty, agreement between the
measurements (including those obtained from exclusive decays) will give us
confidence that both theoretical and experimental errors are under control.

The main theoretical development since the last CKM workshop \cite{Battaglia:2003in}
is the study of $1/m_b$ corrections in ratio of rates such as:
\be
R_{s.l./\gamma}=
\frac{\Gamma_{X_u\ell\nu}(E_\ell\ge E_c)}{\Gamma_{X_s\gamma}(E_\gamma\ge E_c)}
\label{eq:rslovgamma}\ ,\ee
where the cut can be on the lepton energy, as in \eq{eq:rslovgamma}, or on the
hadronic invariant mass. These ratios are important because the
dependence of the individual rates on the universal, non-perturbative shape
function $f(k_+)$, which describes the distribution of the light-cone
component of the residual momentum of the $b$ quark, cancels up to
perturbative and subleading-twist corrections. This allows for a
model-independent determination of $|V_{ub}|$ (rather
$|V_{ub}|/|V_{tb}V_{ts}^*|$) at that order. 

Obviously, the accuracy of these methods depends on the size and
uncertainty of subleading corrections. Perturbative corrections are
dominated by large Sudakov logarithms which were summed to subleading
order some time ago 
\cite{Leibovich:1999xf,Leibovich:2000ey}. $1/m_b$ corrections, on the
other hand, have only been studied recently. There are, in
fact, two large effects.

The first is a higher-twist correction which arises at
$\ord{1/m_b}$ 
\cite{Bauer:2001mh,Bauer:2002yu,Leibovich:2002ys,Neubert:2002yx}. 
For a cut $E_\ell>2.3\,\gev$, models indicate that it leads to an
order 15\% upward shift in the value of $|V_{ub}|$.  Though
substantial, this correction may be insensitive to the model chosen
for the subleading distribution function
\cite{Neubert:2002yx}. Moreover, these effects can be circumvented for
a large part in hadronic-mass-cut measurements \cite{Bigi:2002qq}.

The second effect is due to a weak annihilation (WA) contribution and
arises at relative order $1/m_b^2$
\cite{Bigi:1994bh,Voloshin:2001xi}. This contribution is the first of
an infinite series which re-sums into a subleading distribution
function \cite{Leibovich:2002ys}.  For a cut $E_\ell>2.3\,\gev$, it is
estimated to be roughly 10\% with unknown sign
\cite{Leibovich:2002ys}. Both this and the higher-twist correction 
are significantly reduced when
the cut on $E_\ell$ is lowered below $2.3\,\gev$.

As emphasized by Luke \cite{Luke:2003nu}, experiment itself can be used
to further reduce theoretical errors in the inclusive measurement of
$|V_{ub}|$. Studying the dependence of $|V_{ub}|$ on the lepton cut,
for instance, can help test the size of subleading twist
contributions. Moreover, comparing the value of $|V_{ub}|$ extracted
from inclusive semileptonic decays of charged and neutral $B$ mesons
will give a handle on WA contributions. Of course, more precise
measurements of the photon spectrum in $B\to X_s\gamma$ will improve
the determination of $|V_{ub}|$ from ratios such as the one in
\eq{eq:rslovgamma}. And for $(q^2,\,m_X)$ cuts, better determinations
of $m_b$ from moments of inclusive $B$ decay rates will reduce the
largest source of uncertainty in the corresponding theoretical
expressions.

\subsection{Inclusive $|V_{ub}|$: experiment}

There have been many new results and analyses since the last CKM
workshop, notably from BABAR and BELLE \cite{Sugiyama:2003rh}, as
presented by Sarti
\cite{Sarti:2003mt} and Kakuno \cite{Kakuno:DurCKM03} at this
workshop. Precision has been improved and BELLE has tried new methods.
The overall situation was very nicely reviewed in Muheim's presentation
\cite{Muheim:2003hm}.

\subsubsection{BABAR 2002: endpoint measurement \cite{Luth:2002mq,Sarti:2003mt}}

The measurement is based on $20.6\,\fbinv$ of on-peak and
$2.6\,\fbinv$ of off-peak data. The cut on the charged lepton energy
is $2.3\,\gev\le E_\ell\le 2.6\,\gev$.  They measure a partial
branching fraction $\Delta\mathcal{B}(B\to X_u\ell\nu)=(0.152\pm
0.014\pm0.014)\cdot 10^{-3}$, where the systematics come from, in
order of importance: the estimate of the efficiency; the continuum
background subtraction; the variations in the beam energy; the $B\bar
B$ background modeling.  They use CLEO's determination
\cite{Bornheim:2002du} of the fraction, $f_u$, of the spectrum that
falls into their momentum interval to obtain the full rate and the
PDG 2002 average $B$ lifetime \cite{Hagiwara:2002fs} to
measure $|V_{ub}|$. They find:
\bea
|V_{ub}|&=&(4.43\pm0.29_{exp}\pm 0.25_{HQE}\pm 0.50_{f_u}\nn\\
&&\pm 0.35_{s\gamma})\cdot 10^{-3}
\ ,\eea
where the labelling of errors should be obvious. 

\subsubsection{BABAR 2003: $m_X$ cut with fully reconstructed $B$'s \cite{Sarti:2003mt}}

This analysis is based on about 88 million $B\bar B$ events
($82\,\fbinv$) in which one of the $B$'s is fully reconstructed
through decays of the form $B\to D^{(*)}\mathrm{hadrons}$ while the
semileptonic decay is measured on the opposing $B$ with a cut
$p_\ell>1\,\gev$ and $m_X<1.55\,\gev$, the latter being optimized to
reduce the total error. This allows to reconstruct both the neutrino
and the hadronic system $X$ and to separate charged and neutral $B$
mesons.  It has the advantage of giving a large phase-space acceptance
and a high purity of the sample. To reduce systematics due to
uncertainties in the efficiency, they normalize the signal by the
total semileptonic branching ratio. They use CLEO's determination of
$\bar\Lambda$ and $\lambda_1$ \cite{Bornheim:2002du} to determine their signal selection
efficiency and to extrapolate the partial rate to the full phase
space. Using BABAR's semileptonic branching fraction \cite{Aubert:2002uf} and the
PDG 2002 average $B$ lifetime \cite{Hagiwara:2002fs}, they obtain:
\bea
|V_{ub}|&=&(4.52\pm0.31_{stat}\pm0.27_{syst}\pm0.40_{extrap}\nn\\
&&\pm0.25_{HQE})\cdot 10^{-3}
\ ,
\eea
where again the labelling of errors should be obvious.  This determination is
the most precise one to date.

\subsubsection{BELLE 2003: $m_X$ cut with $B\to D^{(*)}\ell\nu$ tagging 
\cite{Sugiyama:2003rh,Kakuno:DurCKM03}}

This analysis is based on a sample of approximately 84 million $B\bar
B$ pairs or $78.1\,\fbinv$. Though it leads to a two-fold degeneracy
in the decaying $B$ meson direction which results from the presence of
a second neutrino, $B\to D^{(*)}\ell\nu$ tagging improves on the
efficiency of full reconstruction without degrading the $m_X$
resolution. With a cut on the signal lepton momentum $p_\ell>1\,\gev$
and on the hadronic recoil mass $m_X<1.5\,\gev$, BELLE obtains the
preliminary branching ratio, $\mathcal{B}(B\to X_u\ell\nu)=(2.62\pm
0.63_{stat}\pm0.24_{syst}\pm0.39_{extrap.})\cdot 10^{-3}$. This
implies
\bea
|V_{ub}|&=&(5.00\pm0.60_{stat}\pm0.24_{syst}\pm0.39_{extrap.}\nn\\
&&\pm0.36_{HQE})\cdot 10^{-3}
\ .
\eea

\subsubsection{BELLE 2003: $(q^2,\,m_X)$ cut with advanced neutrino 
reconstruction 
\cite{Sugiyama:2003rh,Kakuno:DurCKM03}}

This method is introduced to increase efficiency while avoiding the
degradation in $(q^2,\,m_X)$ resolution brought about by hermiticity
based neutrino reconstruction. Events with only one charged lepton ($e$ or $\mu$)
are retained. The neutrino momentum is then calculated by subtracting the four-momenta
of all reconstructed particles from that of the $\Upsilon(4S)$. This calculation
is improved by reconstructing the other $B$ decay through a simulated annealing technique.
The signal region is then defined by $m_X<1.5\,\gev$ and $q^2>7\,\gev$. The resulting
branching fraction is $\mathcal{B}(B\to X_u\ell\nu)=(1.64\pm
0.14_{stat}\pm0.46_{syst}\pm0.22_{extrap.})\cdot 10^{-3}$, yielding
\bea
|V_{ub}|&=&(3.96\pm0.17_{stat}\pm0.56_{syst}\pm0.26_{extrap.}\nn\\
&&\pm0.29_{HQE})\cdot 10^{-3}
\ .
\eea

\subsubsection{Summary of inclusive $|V_{ub}|$ measurements \cite{Muheim:2003hm}}

\begin{figure}[t]
\includegraphics[width=8cm]{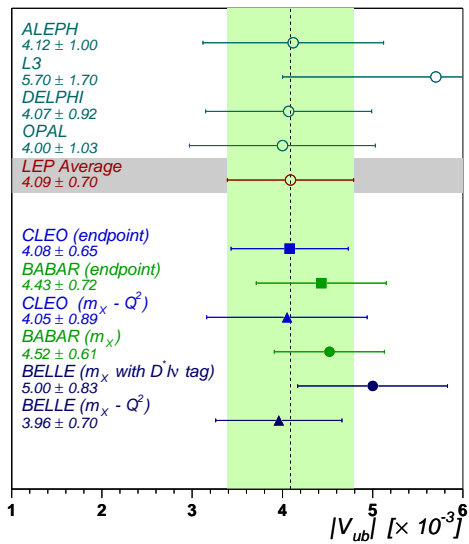}
\caption{\label{fig:vubinclhfag}Compilation of inclusive $|V_{ub}|$ results by the 
Heavy Flavor Averaging Group (HFAG) for the 2003 winter conferences \cite{cite:HFAG}.}
\end{figure}

Muheim combined the above results with the CLEO and BABAR endpoint measurements
as well as CLEO's $(q^2,\,m_X)$ determination to obtain the following $B$ factory value for
$|V_{ub}|$ from inclusive $B$ decays:
\be
|V_{ub}|=(4.32\pm0.57)\cdot 10^{-3}
\ ,
\ee
which improves on the LEP average: $|V_{ub}|=(4.09\pm0.70)\cdot 10^{-3}$. All of the
relevant measurements are summarized in \fig{fig:vubinclhfag}.

\section{Moments of inclusive $B$ decay spectra and $|V_{cb}|$}
\label{sec:moments}

Since different moments depend differently on the various parameters
or non-perturbative quantities which appear in the HQE that describe
inclusive $B$ decays, a measurement of moments allows a determination
of these quantities and numerous consistency checks of the HQE. One
should not forget, however, that there is an assumption behind these
determinations: in order to be sensitive to non-perturbative
$1/m_b$ corrections which are formally smaller than any term in
perturbation theory, one must assume that the former are larger than
the perturbative terms neglected.

The moments which have been considered most frequently up to now are
moments of the photon energy spectrum in $B\to X_s\gamma$ decays,
moments of the charged lepton energy spectrum and of the hadronic
recoil mass in $B\to X_c\ell\nu$ decays. Moments analyses were
pioneered by CLEO \cite{Cronin-Hennessy:2001fk}. Today, a sufficient
variety of moments have been measured to allow a global fit to the
corresponding HQE expressions up to and including $1/m_b^3$ terms
\cite{Bauer:2002sh,Battaglia:2002tm}.

Experimental aspects of the subject were reviewed very nicely by Calvi
\cite{Calvi:2003vt} at this workshop and Luke \cite{Luke:2003nu} and
Uraltsev \cite{Uraltsev:2003bw} provided very interesting discussions
of some of the theoretical issues involved. Results from BABAR were
presented by Luth \cite{Luth:2003mr} and from CLEO by Cassel
\cite{Cassel:2003ru}. The subject was also covered quite extensively
in the proceedings of the 1st CKM workshop \cite{Battaglia:2003in}.

The main novelty since the last workshop are the global fits mentioned
above, whose results were actually summarized in the proceedings
\cite{Battaglia:2003in}.  In
\cite{Battaglia:2002tm} the authors use preliminary 
DELPHI measurements of the first
three moments of the hadronic mass and charge lepton energy spectra
\cite{Calvi:2002wc} and obtain
\bea
m_b(1\,\gev)&=&4.59\pm0.08\,\gev\nn\\
m_c(1\,\gev)&=&1.13\pm0.13\,\gev\nn\\
|V_{cb}|&=&(41.1\pm1.1)\cdot 10^{-3}\label{eq:delphiinclvcb}
\ ,\eea
where no expansion in $1/m_c$ is performed and matrix elements up to
$\ord{1/m_b^3}$ are also obtained. The masses given here are the
running kinetic masses. The corresponding $1S$ mass for the $b$ is
$m_b^{1S}=4.69\pm0.08\,\gev$. The quality of their fit indicates the
consistency of the HQE description of these moments at the order
considered. In \cite{Bauer:2002sh}, the authors use a total of 14
moment measurements from CLEO \cite{Briere:2002hw}, BABAR
\cite{Aubert:2002pm} and DELPHI
\cite{Calvi:2002wc}. Imposing the constraint on $m_b-m_c$ given by 
the $B^{(*)}$ and $D^{(*)}$ masses, 
thereby introducing an $1/m_c$ expansion, they obtain
\bea
m_b^{1S}&=&4.74\pm0.10\,\gev\nn\\
|V_{cb}|&=&(40.8\pm0.9)\cdot 10^{-3}\label{eq:allinclvcb}
\ ,\eea
along with matrix elements up to $\ord{1/m_b^3}$. The 2-3\% accuracy
of these $|V_{cb}|$ measurements are impressive. They are currently
limited by the accuracy of the moments measurements and should
therefore be improved in the near future with additional data from the
$B$ factories.

\begin{figure}
\includegraphics*[width=8cm]{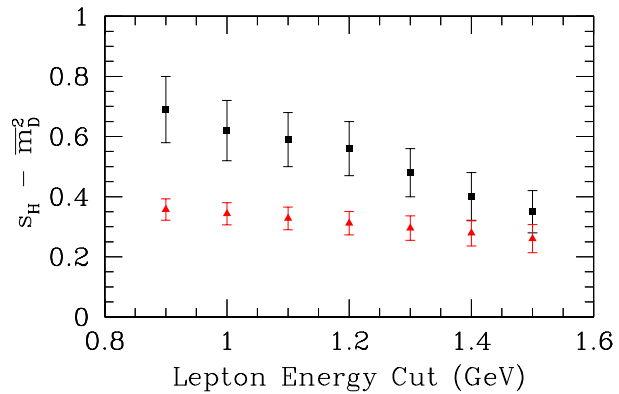}%
\caption{\label{fig:mx2vselcut} Comparison of BABAR's measurement of the 
moment of the hadron invariant mass spectrum \cite{Luth:2003mr}
vs. lepton energy cut (squares) with the HQE expansion obtained
\cite{Bauer:2002sh} by fixing its parameters with CLEO's
$E_\ell^{cut}=1.5\,\gev$, $\la m_X^2\ra$ \cite{Cronin-Hennessy:2001fk}
and $B\to X_s\gamma$, $\la E_\gamma\ra$ measurements
\cite{Chen:2001fj}.}
\end{figure}

It is important to note that the results of \cite{Bauer:2002sh}
include the $m_X^2$ moment measurement of BABAR presented at
this workshop by Luth
\cite{Luth:2003mr}, whose dependence on the charge lepton energy cut 
appears to
be in contradiction with the HQE. This measurement is based on a
sample of 55 million $B\bar B$ pairs. In these events, one of the
$B$'s is fully reconstructed while the semileptonic decay of the
second is identified by a high momentum charged lepton. The
discrepancy arises when the HQE parameters are fixed at
$E_\ell^{cut}=1.5\,\gev$ and the predicted moment is compared with
data for lower values of $E_\ell^{cut}$, as shown in
\fig{fig:mx2vselcut}.

A possible resolution was proposed by Luke and collaborators
\cite{Bauer:2002sh}.  The measurement depends on the assumed spectrum
of excited $D$ resonances, in which there are no contributions from
excited states with masses below $\sim 2.4\,\gev$. The addition of a
non-negligible fraction of excited $D$ states with masses less than
$2.45\,\gev$ could help reconcile the discrepancy.

Another solution to this problem was put forward by Uraltsev
\cite{Uraltsev:2003bw}. He emphasized that the convergence of the HQE 
is governed by
the maximum energy release or hardness of the moment considered. For
$E_\ell=1.5\,\gev$, the hardness is only $1.25\,\gev$ and smaller than
$1\,\gev$ for $E_\ell>1.7\,\gev$ , implying rather poor convergence in
this region of phase space. And the hardness only decreases for higher
moments. Thus, the problem with the BABAR moment measurement is not at
low $E_\ell$, but rather at high $E_\ell$, where the matching to the
HQE is actually performed.  His recommendation is therefore to perform
comparisons to the HQE at the lowest practical value of $E_\ell$.

The main message of this discussion is that experimental groups should
strive as much as possible to obtain model-independent measurements of
these spectral moments and the applicability of the HQE in dangerous
regions of phase space should be considered with care.

Since this discussion took place, new measurements of
hadronic moments as a function of lepton-energy cut have been
presented by BABAR \cite{Aubert:2003dr} and CLEO
\cite{Huang:2003ay}. Both these measurements are preliminary. 
CLEO derives the $m_X^2$ moment for a number of $E_\ell^{cut}$ in the
range of 1 to 1.5 GeV, from the branching fractions and average hadron
mass distributions of a number of charm meson resonant and
non-resonant states, as advocated in
\cite{Cronin-Hennessy:2001fk}. They obtain,
\bea
\la m_X^2-m_D^2\ra_{E_\ell>1.0\,\gev}&=&0.456\pm0.014_{stat}\nn\\
&&\pm0.045_{detect}\nn\\
&&\pm0.109_{model}\,\gev\\
\la m_X^2-m_D^2\ra_{E_\ell>1.5\,\gev}&=&0.293\pm0.012_{stat}\nn\\
&&\pm0.033_{detect}\nn\\
&&\pm 0.048_{model}\,\gev
\label{eq:cleomx215}\ .\eea
BABAR has inaugurated a new method in which the $m_X$ and $m_X^2$
moments are extracted directly from the measured $m_X$ and $m_X^2$
distributions. This analysis reduces dependence on the mass
distributions and branching fractions of individual charm states which
are poorly known for higher mass states. Combining their results
for $\la m_X^2\ra$ with their earlier measurements of semileptonic
branching ratios and $B$ lifetimes, they obtain
\bea
m_b^{1S}&=&4.638\pm0.094_{expt}\pm0.090_{thy}\,\gev\nn\\
|V_{cb}|&=&(42.10\pm 1.04_{expt}\pm0.72_{thy})\cdot 10^{-3}
\label{eq:babarinclvcb}
\ ,\eea
in good agreement with the results of
\eqs{eq:delphiinclvcb}{eq:allinclvcb}. 
They also find $\lambda_1=-0.26\pm 0.06_{expt}\pm0.06_{thy}\,\gev^2$.

\begin{figure}
\includegraphics*[width=8cm]{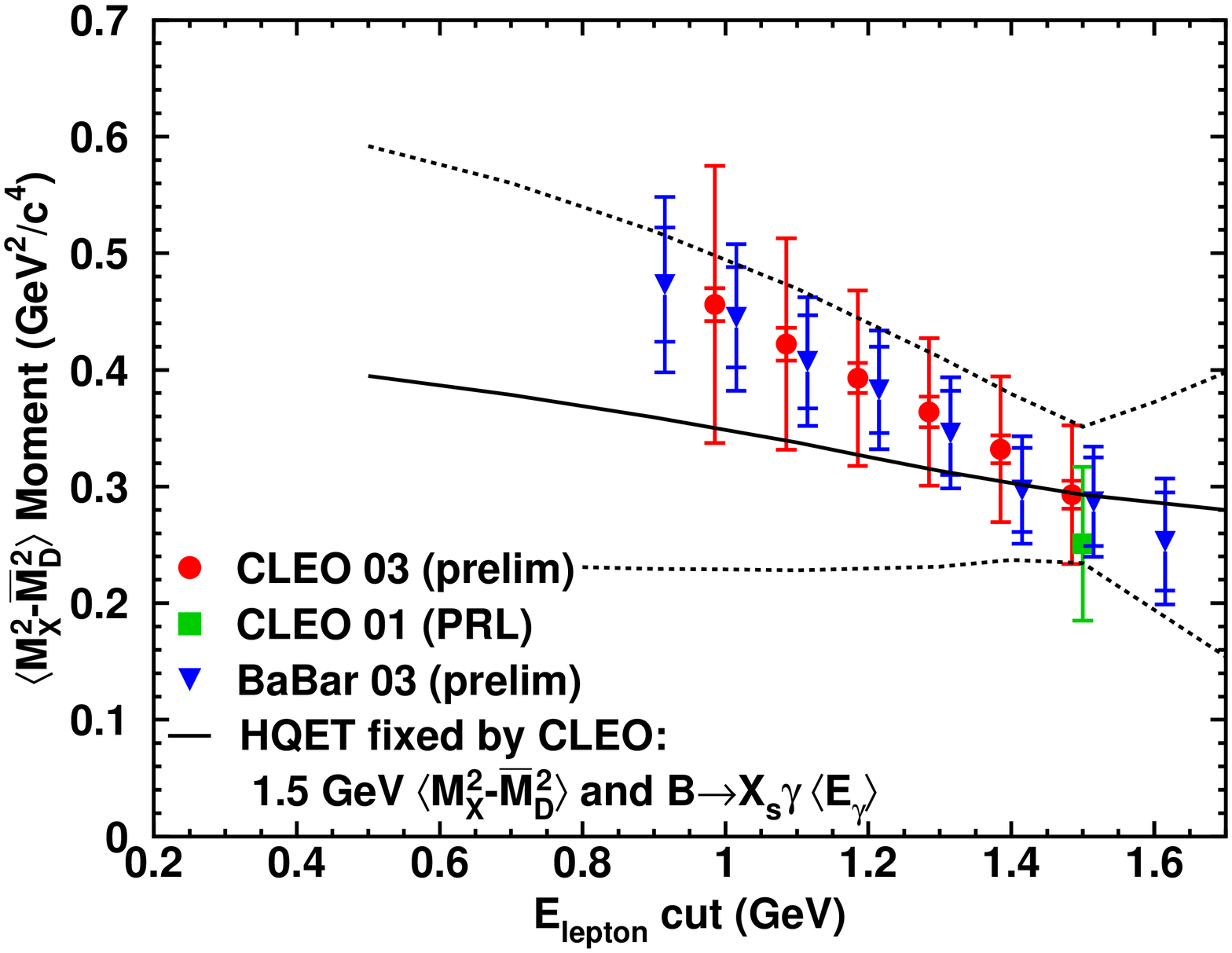}
\caption{\label{fig:mx2vselcutnew} 
Comparison of BABAR '03 \cite{Aubert:2003dr}, 
CLEO '03 \cite{Huang:2003ay} and CLEO '01 \cite{Cronin-Hennessy:2001fk} 
measurements of the 
moment of the hadron invariant mass spectrum 
vs. lepton energy cut. 
The theory bands shown in 
the figure reflect the variation of the experimental errors on the two
constraints, the variation of the third-order HQET parameters by the
scale $(0.5\,\gev)^3$, and variation of the size of the higher
order QCD radiative corrections \cite{Bauer:2002sh}. Figure taken
from \cite{Huang:2003ay}.}
\end{figure}

As a result of several changes to their analysis and data selection,
BABAR find that their new results for $\la m_X^2\ra$
vs. $E_\ell^{cut}$ fall substantially below those reported in
\cite{Luth:2003mr} and depicted in \fig{fig:mx2vselcut} at low
$E_\ell^{cut}$. This has the effect of reconciling experiment with
theory as shown in
\fig{fig:mx2vselcutnew}, where BABAR and CLEO's results are plotted together
with the theoretical prediction constrained by CLEO's measurement at
$E_\ell^{cut}=1.5\,\gev$ (\eq{eq:cleomx215}) and the first $B\to
X_s\gamma$ photon energy moment \cite{Chen:2001fj}. Agreement
between the two experiments is excellent.

\section{$|V_{ub}|$ from exclusive semileptonic $B$ decays}
\label{sec:exclusiveVub}

Many new measurements of exclusive $b\to u\ell\nu$ decays using new
techniques have been presented recently, with more to come. These were
very nicely reviewed by Gibbons/Cassel \cite{Gibbons:2003gq}, with
presentations from BABAR, BELLE and CLEO by Schubert
\cite{Aubert:2003zd}, Schwanda \cite{Schwanda:2003bj} and
Gibbons/Cassel \cite{Gibbons:2003gq}. The improving statistics of
experiments are beginning to permit the measurement of partial rates
as a function of lepton recoil squared, $q^2$, allowing reduction of the
dependence of the measured rates and $|V_{ub}|$ on the still rather
poorly known theoretical form factor shapes. Such measurements also
help eliminate incorrect form factor models. The overall
normalizations of the form factors, however, cannot be tested
experimentally and dominate the extraction of $|V_{ub}|$ from measured
rates. One therefore needs model-independent determinations of these
form factors such as those which upcoming, unquenched lattice QCD
calculations should provide.

\subsection{Exclusive $|V_{ub}|$: theory}

The theory of exclusive, semileptonic $b\to u\ell\nu$ has evolved
little since the publication of the proceedings from the last CKM
workshop \cite{Battaglia:2003in}. The status of lattice QCD (LQCD) calculations of
$B^0\to\pi^-(\rho^-)\ell^+\nu$ form factors was very nicely reviewed
by Onogi \cite{Onogi:2003kf} and that of light-cone sum-rule (LCSR) calculations by Ball
\cite{Ball:2003rd}. One important feature of lattice calculations is that they
are currently limited to smaller recoils ($q^2\gsim10\,\gev^2$) while LCSR
calculations are more reliable at larger recoils ($q^2\lsim 15\,\gev^2$).

\begin{figure}[t]
\begin{center}
\includegraphics[width=8cm]{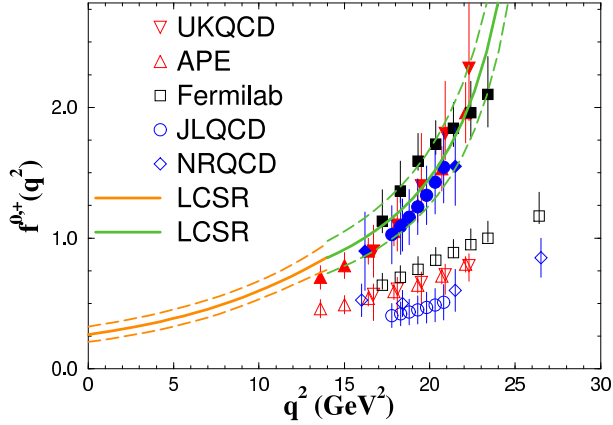}
\end{center}
\caption{\label{fig:bpiff}
$B^0\to\pi^-\ell^+\nu$ form factors from different lattice groups
(UKQCD \cite{Bowler:1999xn}, APE \cite{Abada:2000ty}, Fermilab
\cite{El-Khadra:2001rv}, JLQCD \cite{Aoki:2001rd}, 
NRQCD \cite{Shigemitsu:2002wh})and from LCSR
\cite{Ball:2001fp}. The LCSR results for $q^2\ge 14\,\gev^2$ are
obtained using a pole ansatz whose residue is fixed by a LCSR
calculation. Figure taken from \cite{Onogi:2003kf}.}
\end{figure}

The current situation regarding quenched lattice calculations of
$B^0\to\pi^-$ form factors is summarized in \fig{fig:bpiff}.  There is
good agreement amongst the different methods used to obtain $f^+(q^2)$, which
determines the rate in the limit $m_\ell=0$, with errors at the level
of 15-20\%. Agreement is also good with the recent LCSR results of
\cite{Ball:2001fp} (see also \cite{Khodjamirian:2000ds}). 
Agreement is less clear for the lattice results for $f^0(q^2)$, due to
sensitivity of this form factor on light and heavy quark masses.

\medskip

\begin{figure}[t]
\begin{center}
\includegraphics[width=8cm]{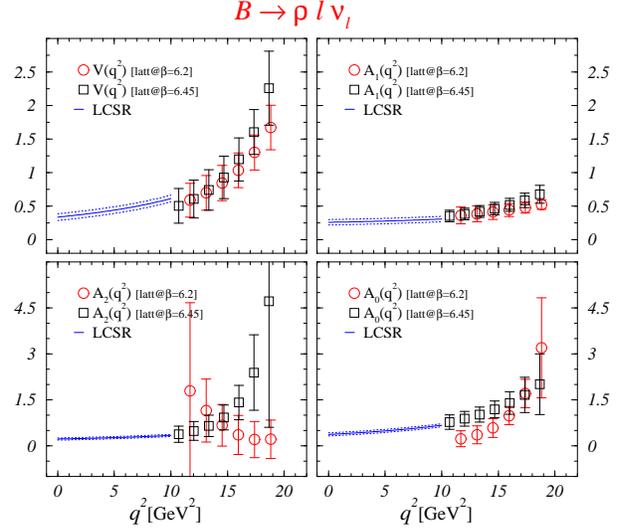}
\end{center}
\caption{\label{fig:brhoff}
Example of quenched lattice results for $B^0\to\rho^- \ell^+\nu$ 
form factors plotted as a function of $q^2$ \cite{Abada:2002ie}. These
results were obtained at two values of the inverse lattice spacing 
$1/a=3.7\,\gev$ and $2.7\,\gev$, corresponding to bare couplings
values $\beta=6.45$ and $6.2$ respectively. Also shown at low
$q^2$ are the light-cone sum rule results of \cite{Ball:1998kk}. }
\end{figure}

The situation for $B^0\to\rho^-\ell^+\nu$ decays is quite
different. There are far fewer calculations of $B\to\rho$ form
factors, both with LQCD and LCSR. Moreover, quenching effects may be
more important here than in $B\to\pi$ decays because the $\rho$ cannot
decay into two $\pi$ in the quenched theory.  Nevertheless, quenched
lattice calculations do provide a first estimate of the relevant
matrix elements which is worth considering. While there are a number
of older lattice calculations
\cite{Abada:1994dh,Allton:1995ui,Flynn:1996dc}, for clarity we only
show in 
\fig{fig:brhoff} the recent, preliminary results of the SPQcdR collaboration
\cite{Abada:2002ie}, obtained at two values of the lattice
spacing. The small dependence on lattice spacing of $A_1$, which
dominates the rate at large $q^2$, indicates that discretization
errors on this form factor are small. Similar results have been
obtained recently by the UKQCD collaboration
\cite{Gill:2001jp}.

Also shown in \fig{fig:brhoff} are the LCSR results of
\cite{Ball:1998kk}. These results look like a rather natural extension
of the lattice results to smaller $q^2$, suggesting rather good
agreement between the two methods.

\begin{figure}[t]
\begin{center}
\includegraphics[width=7cm]{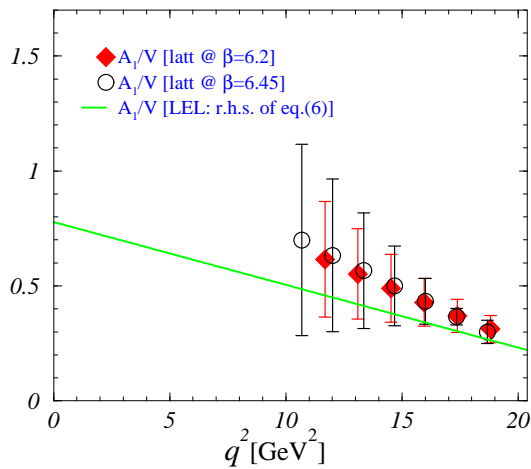}
\end{center}
\caption{\label{fig:brhoscet}
Data points are $A_1(q^2)/V(q^2)$ vs. $q^2$ from the lattice calculation
of \cite{Abada:2002ie}.
The solid line corresponds to the r.h.s. of \eq{eq:a1ovvscet}.}
\end{figure}

Another interesting feature of the lattice $B\to\rho$ calculations is
the agreement with SCET constraints such as
\cite{Charles:1998dr,Bauer:2000yr,Burdman:2000ku,Beneke:2002ph}:
\be
\frac{A_1(q^2)}{V(q^2)}=\frac{2E_\rho M_B}{(M_B+m_\rho)^2}
\label{eq:a1ovvscet}
\ ,
\ee
as shown in \fig{fig:brhoscet}.

\medskip

To extend lattice results to smaller values of $q^2$ in a
model-independent way one can make use of dispersive bounds
\cite{Boyd:1995tt,Lellouch:1996yv}. While there are ways of improving
these bounds, for the moment they do not provide sufficient accuracy.
One may therefore wish to consider a combination of LCSR results at
low $q^2$ and quenched LQCD results at high $q^2$. This approach
offers a reasonably reliable determination of the form factors over
the full kinematic range and has already been used for $|V_{ub}|$
determinations \cite{Athar:2003yg}. To avoid the problem
``extrapolating'' lattice results to lower values of $q^2$ altogether,
one can also consider extractions of $|V_{ub}|$ from the partial rates
measured for $q^2\gsim 12\,\gev^2$, as was suggested in
\cite{Flynn:1996dc} and already implemented in \cite{Athar:2003yg}.

As they stand, quenched LQCD and LCSR results have errors of order
20\%, which is not sufficient given the quality of the experimental
measurements to come. While significant improvement of LCSR results
cannot be expected, lattice predictions can and will be
improved. Other than the issue of extending lattice results to smaller
values of $q^2$, one of the main issues in these calculations is that
of quenching.  Indeed, only fully unquenched lattice calculations will
provide completely model-independent determinations of the relevant
form factors. Partially unquenched results with two flavors of Wilson
sea quarks are expected soon from JLQCD and UKQCD and three
Kogut-Susskind (KS) flavor calculations based on MILC configurations
should also be forthcoming. While JLQCD and UKQCD will be limited to
light quark masses $\gsim m_s/2$, the MILC configurations extend down
to $\sim m_s/8$. This means that the uncertainties associated with the
necessary extrapolations to the physical $u$ and $d$ quark masses
should be much smaller in the calculations performed on these
configurations. On the other hand, the methods used to produce the
MILC configurations may introduce non-localities and KS fermions
suffer from flavor violations which can be accounted for but which
significantly complicate chiral extrapolations.

Another important avenue to explore to reduce errors in LQCD
calculations are ratios of semileptonic $B$ and $D$ meson
rates, as many systematic and statistical errors are expected to cancel in
such ratios. Results for $B$ mesons can then be recovered by combining
these lattice ratios with the high-precision measurements of $D$
decays promised by CLEO-c.  For a more complete discussion of both
LQCD and LCSR calculations, please see the CKM workshop yellow book
\cite{Battaglia:2003in} and the reviews by Onogi
\cite{Onogi:2003kf} and by Ball \cite{Ball:2003rd}.

\subsection{Exclusive $|V_{ub}|$:  experiment}

As already mentioned, the last year has seen many new measurements of
exclusive $b\to u\ell\nu$ decays, many of which are still preliminary. All of these
measurements make use of detector hermiticity to reconstruct the four-momentum
of the neutrino. They are:
\begin{itemize}
\item
BABAR 2003 \cite{Aubert:2003zd}, reported at this workshop by Schubert: measurement of 
$B\to\rho\ell\nu$ rate with an on resonance integrated luminosity of $L_{on}=50.5\,\fbinv$,
an off resonance luminosity of $L_{off}=7.8\,\fbinv$ and the following cut on the lepton
momentum: $2.0\,\gev<p_\ell<2.7\,\gev$. 

\item
BELLE 2003 (preliminary), reported at this workshop by Schwanda
\cite{Schwanda:2003bj}: first measurement of the $B\to\omega\ell\nu$ rate with
$L_{on}=78.1\,\fbinv$, $L_{off}=8.8\,\fbinv$ and
$1.5\,\gev<p_\ell<2.7\,\gev$. They find
$\mathcal{B}(B^+\to\omega\ell^+\nu)=(1.3\pm0.4_{stat}\pm0.2_{syst}\pm0.3_{model})\cdot
10^{-4}$.

\item
BELLE 2002 (preliminary) \cite{Kwon:ICHEP02}: 
\begin{itemize}
\item
measurement of $B\to\pi\ell\nu$ rate with $L_{on}=60\,\fbinv$, $L_{off}=9\,\fbinv$ and 
$1.2\,\gev<p_\ell<2.8\,\gev$, including a measurement of the differential decay rate
as a function of $q^2$.
\item
measurement of $B\to\rho\ell\nu$ rate with $L_{on}=29\,\fbinv$, $L_{off}=3\,\fbinv$ and 
$2.0\,\gev<p_\ell<2.8\,\gev$.
\end{itemize}

\item
CLEO 2003, reported at this workshop by Gibbons/Cassel \cite{Gibbons:2003gq}:
measurements of $B\to\pi\ell\nu$ and $B\to\rho\ell\nu$ rates based on
a sample of 9.7 million $B\bar B$ pairs with $p_\ell>1.0\,\gev$ for
pseudoscalar final states and $p_\ell>1.5\,\gev$ for vector final
states. This study pioneers a new method in which rates are measured
independently in three $q^2$ bins, yielding reduced
model-dependence and allowing for model discrimination.

\end{itemize}

\begin{figure}
\includegraphics[width=8cm]{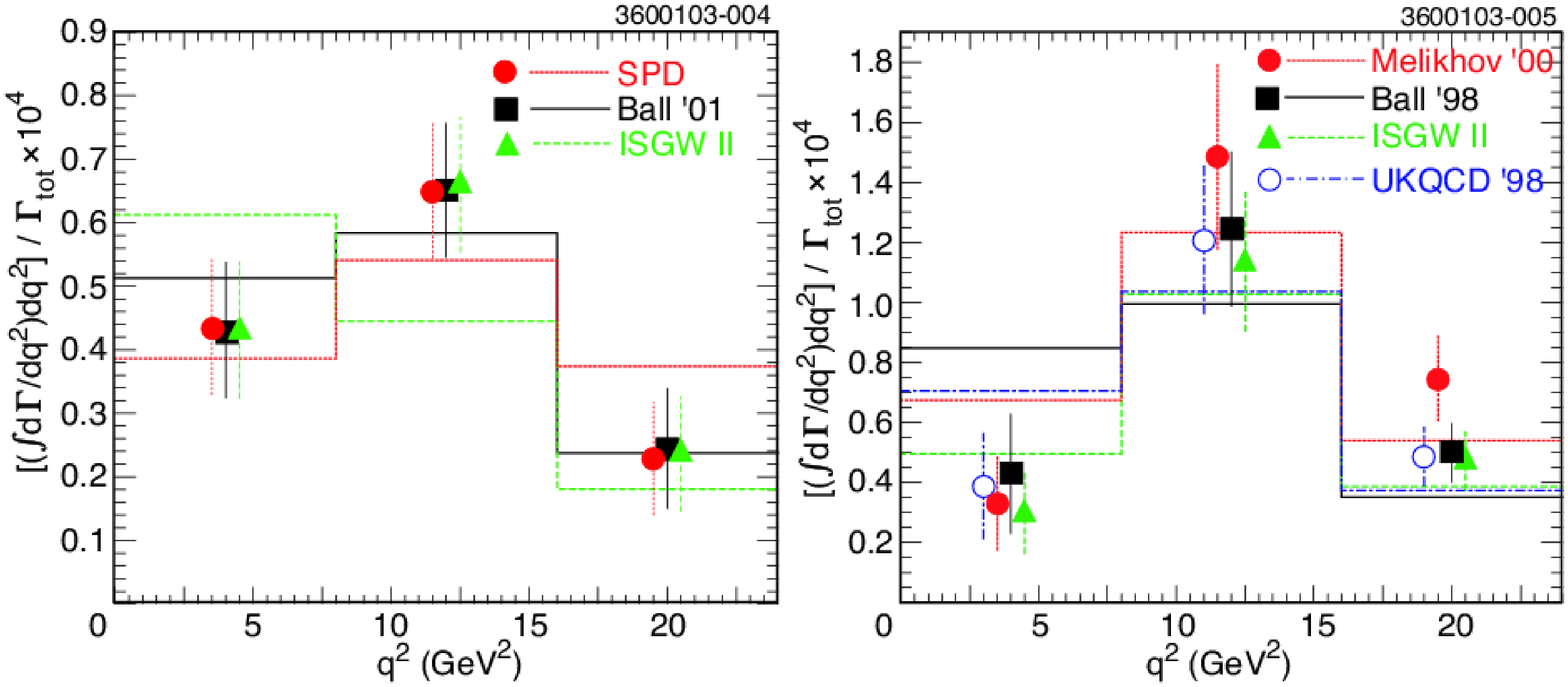}
\caption{\label{fig:bpibrhovsq2cleo}
The $d\Gamma/dq^2$ distributions obtained in the CLEO '03 analysis for 
$B^0\to\pi^-\ell^+\nu$ (left) and $B^0\to\rho^-\ell^+\nu$ (right).  Shown are the variations in 
the extracted 
rates (points) for form factor calculations that have significant $q^2$ variations, and the best fit of 
those shapes to the extracted rates (histograms). Plot taken from \cite{Gibbons:2003gq}.}
\end{figure}

CLEO's results for the $q^2$ dependence of the partial rates for
$B^0\to\pi^-\ell^+\nu$ and $B^0\to\rho^-\ell^+\nu$ are shown in
\fig{fig:bpibrhovsq2cleo}, as obtained using different form factor
calculations to estimate efficiencies. The results for
$B^0\to\pi^-\ell^+\nu$ show negligible dependence on the calculation
used, indicating that their binning method has essentially eliminated
form factor dependence. The situation is less good for
$B^0\to\rho^-\ell^+\nu$ decays, likely a result of the cut on the
angle between the lepton and the $W$ directions
\cite{Gibbons:2003gq}. The poor $\chi^2$ for the ISGWII model
\cite{Scora:1995ty} fit to the $B^0\to\pi^-\ell^+\nu$ rate and for the
Melikhov {\it et al.} model \cite{Melikhov:2000yu} and Ball {\it et
al.}  LCSR \cite{Ball:1998kk} fit to the $B^0\to\rho^-\ell^+\nu$ rate
indicate that these theoretical descriptions of the form factors are
disfavored by the data.

BELLE also has a determination of $d\Gamma/dq^2(B^0\to\pi^-\ell^+\nu)$
as a function of $q^2$ as shown in \cite{Kwon:ICHEP02}. Unlike CLEO,
BELLE determines its efficiency without binning in $q^2$. The
model-dependence of their result is therefore expected to be more
important, though it has not yet been determined.

\medskip

\begin{figure}
\includegraphics[width=8cm]{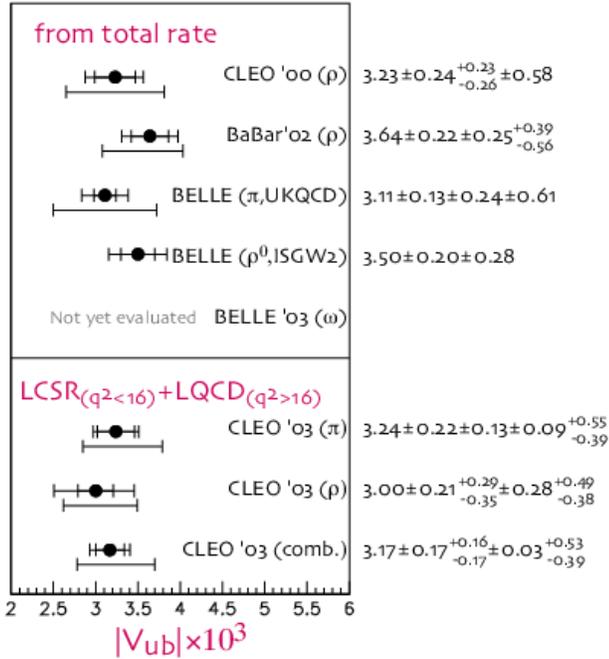}
\caption{\label{fig:vubsummary}
Compilation $|V_{ub}|$ measurement from exclusive $B\to
X_u\ell\nu$ decays \cite{Gibbons:2003gq}. }
\end{figure}

A compilation of results for $|V_{ub}|$ obtained from exclusive $B\to
X_u\ell\nu$ decays is shown in \fig{fig:vubsummary}. Gibbons 
refrains from giving an average number because the current
information provided by the different experiments is insufficient to
determine the size of correlations in their results.
Indeed, there are a number of common systematics which could lead
to large correlations \cite{Gibbons:2003gq}. These include:
\begin{itemize}
\item
the common use of the ISGW2 model \cite{Scora:1995ty} and the
Neubert-Fazio model \cite{DeFazio:1999sv} to determine the $b\to
u\ell\nu$ background coming from feed down modes not considered;
\item
common GEANT base for detector simulation;
\item
common signal models: LCSR, LQCD, quark models.
\end{itemize}
The first item should probably be treated as a correlated systematic.
To deal with the issue of common signal models, it seems appropriate
to first average the rates and $|V_{ub}|$ obtained by the different
experiments for a given model and then combine the measurements.

Nevertheless, the good agreement between the different 
experiments in their measurements of $|V_{ub}|$ and in the branching
ratios from which these measurements were obtained is
encouraging. It should be noted, however, that all of these $|V_{ub}|$
determinations are systematically below those obtained from inclusive
decays.

\medskip

With the growing data sets from the $B$ factories, fully reconstructed
$B$-tag analyses, such as those used in the study of inclusive $B\to
X_u\ell\nu$ decays, will become possible. This will reduce background
significantly, allowing for selection criteria which yield a more
uniform efficiency. Consequently, systematic uncertainties associated
with form factor uncertainties and detector and background modeling
will be reduced. At that point, measurement of exclusive $B\to
X_u\ell\nu$ decays may yield the most accurate determinations of
$|V_{ub}|$.

\subsection{Exclusive $|V_{ub}|$ from $\bar B^0\to X_u^+D_s^-$}

At the workshop, Mikami from BELLE \cite{Mikami:DurCKM03} suggested
measuring $|V_{ub}|$ from wrong charm exclusive $\bar B^0\to
\pi^+D_s^-$ decays \cite{Hayakawa:2002ss} and semi-inclusive $\bar
B^0\to X_u^+ D_s^-$ decays \cite{Aleksan:1999iy}, and presented
measurements for the relevant rates and yield.  These decays occur
through the tree-level $b\to u\bar cs$ diagram. The corresponding
measurements of $|V_{ub}|$ are meant as consistency checks for the
usual semileptonic determinations and for the methods used in the
calculation of non-leptonic $B$ decays.

In the semi-inclusive case, the idea would be to obtain
$|V_{ub}/V_{cb}|$ from the endpoint of the $D_s$ spectrum in $\bar
B^0\to X^+ D_s^-$. The advantage with respect to the semileptonic
endpoint measurement is that the signal fraction is larger, with more
than 50\% of the spectrum for $\bar B^0\to X_u^+D_s^-$ beyond the
kinematic limit for $\bar B^0\to X_c^+D_s^-$
\cite{Aleksan:1999iy}. These semi-inclusive decays also have higher
statistics than the exclusive mode.

The problem with both the exclusive and semi-inclusive proposals for
determining $|V_{ub}|$ is that the theoretical formalism to describe
the corresponding decays does not yet exist. Indeed, BBNS
factorization \cite{Beneke:2000ry} does not apply to $B^0\to
\pi^-D_s^+$, because the $\pi^-$ contains the spectator $d$ quark. The
situation is even more complicated for the semi-inclusive case.

More promising, at least theoretically, is the fully inclusive $b\to
u\bar c s'$, as first proposed in \cite{Beneke:1997hv}. In
\cite{Falk:1999sa}, it is shown that when the rate is normalized by
the inclusive semileptonic $b\to c$ rate, the corresponding
theoretical expressions have a well a behaved HQE. This is, of course,
a very challenging measurement to make. However, because it is so
theoretically clean, it would be interesting to investigate
experimental feasibility.

\section{$|V_{cb}|$ from exclusive decays}
\label{sec:exclusiveVcb}

Experimental aspects of these determinations were reviewed very nicely
at this workshop by Oyanguren \cite{Oyanguren:2003vs} and lattice
results for the relevant decay form factors were very nicely
summarized by Onogi \cite{Onogi:2003kf}. The status of these
measurements, as well as the theory behind them, has not evolved
significantly since the publishing of the CKM workshop yellow book
\cite{Battaglia:2003in}. There are no new measurement and no new calculations
of $F(1)$ and $G(1)$, the values of the $B\to D^*\ell\nu$ and $B\to
D\ell\nu$ form factors at zero recoil. Thus, $F(1)=0.91(4)$ and
$G(1)=1.04(6)$
\cite{Battaglia:2003in}. And the extrapolation of the measured rate to
the zero-recoil point, which is required to obtain $|V_{cb}|$, is
still best done with the model-independent, dispersive
parameterizations of \cite{Boyd:1997kz,Caprini:1998mu}, which are given
in terms of a single parameter: the slopes $\rho_{D^*}$ and $\rho_{D}$
of the relevant form factors.

The Heavy Flavor Averaging Group has averaged the results for
$|V_{cb}|$ and the slopes $\rho_{D^*}$ and $\rho_{D}$ obtained by the
different experiments, after rescaling them to common input
\cite{cite:HFAG}. They find:
\bea
|V_{cb}|&=&(42.6\pm 0.6_{stat}\pm 1.0_{syst}\pm 2.1_{thy})\cdot 10^{-3}\nn\\
\rho_{D^*}&=&1.49\pm 0.05_{stat}\pm 0.14_{syst}\nn
\eea
from $B\to D^*\ell\nu$ decays and
\bea
|V_{cb}|&=&(40.8\pm 3.6_{expt}\pm 2.3_{thy})\cdot 10^{-3}\nn\\
\rho_{D}&=&1.14\pm 0.16_{expt}\nn
\eea
for $B\to D\ell\nu$ decays. These measurements are in good agreement
with each other as well as with those obtained using inclusive $B\to
X_c\ell\nu$ decays (e.g. \eqs{eq:delphiinclvcb}{eq:allinclvcb}),
though errors on the exclusive measurements are currently
larger. These errors are dominated by the uncertainties in the
theoretical determination of the form factors at zero recoil. It is
thus important that the quenched lattice calculations \cite{Hashimoto:2001nb}, which
enter the determination of these form factors as
explained in \cite{Battaglia:2003in}, be repeated by other groups and be
unquenched. On the experimental side, the limiting systematics are
inputs such as the $b\to \bar B^0$ and $\Upsilon(4S)\bar B^0$ rates, the
contributions of the $D^{**}$ and the $D$ decay branching ratios.

It should be noted that a new set of very interesting lower bounds
has been derived for the moduli of the derivatives of the Isgur-Wise
function, $\xi(w)$, as explained by Oliver at this workshop
\cite{LeYaouanc:2003xf}.  They are based on derivatives of non-zero
recoil sum rules à la Uraltsev
\cite{Uraltsev:2000ce}. In particular, it is shown that the $n$-th derivative at zero
recoil, $\xi^{(n)}(1)$, can be bounded by the $(n-1)$-st one and that
one obtains an absolute lower bound on the $n$-th derivative,
$(-1)^n\xi^{(n)}(1)\ge (2n+1)!!/2^{2n}$.  Moreover, these bounds are
compatible with the dispersive parameterizations of
\cite{Boyd:1997kz,Caprini:1998mu} and reduce the allowed range of parameters, though it should
be noted that the latter include finite mass corrections which are
absent in the new bounds. It would be interesting to investigate how
radiative corrections and subleading corrections in powers of $1/m_c$
affect these bounds.

\section{$b$-hadron lifetimes and lifetime differences}
\label{sec:lifetimes}

The current experimental situation for $b$-hadron lifetimes and
lifetime differences was very nicely reviewed by Rademacker at this
workshop \cite{Rademacker:2003ux}. The status of both experiment and
theory has not changed significantly since the publication of the CKM Workshop yellow
book
\cite{Battaglia:2003in}. 

\subsection{Lifetimes}

On the theory side, there are no new results and
regarding experiment, the halving of errors
on $\tau_{B^+}$ and $\tau_{B^0}$ brought about by the measurements of
the $B$ factories based on 1999-2001 data were already taken into
account in \cite{Battaglia:2003in}. There are, nevertheless, new measurements from
BABAR \cite{Aubert:2002ms,Aubert:2002sh,deRe:talk02034}, CDF 
\cite{CDF:lifetimesEPS03} and DELPHI \cite{DELPHI:lifetimesEPS03} for both
$\tau_{B^0}$ and $\tau_{B^+}$ and from D0 for $\tau_{B^+}$ \cite{Podesta:March03}. CDF has also 
reported new determinations of $\tau_{B_s}$ and $\tau_{\Lambda_b}$
\cite{CDF:lifetimesEPS03}. These results
are summarized in \tab{tab:newtaub}. The new BABAR measurements are
based on partial reconstruction of the $B$ mesons, either through
$B^0\to D^{*-}(\pi^+,\rho^+)$, where only the $D^0$ in $D^{*-}\to
D^0\pi^-$ is reconstructed \cite{Aubert:2002ms}; through $B^0\to
D^{*-}\ell^+\nu$ \cite{Aubert:2002sh}; and in the di-lepton channel
\cite{deRe:talk02034}. Partial reconstruction works thanks to the
decay kinematics at the $\Upsilon(4S)$.

\begin{table*}[t]
\begin{center}
\begin{tabular}{llc}
\hline
\hline
\multicolumn{3}{c}{New $b$-hadron lifetime measurements}\\
\hline
BABAR \cite{Aubert:2002ms} & $B^0\to D^{*-}\mathrm{(partial)}(\pi^+,\rho^+)$ & 
$\tau_{B^0}=1.553(34)(38)\,\ps$\\
BABAR \cite{Aubert:2002sh} & $B^0\to D^{*-}\ell^+\nu$ & 
$\tau_{B^0}=1.523^{+24}_{-23}(22)\,\ps$\\
BABAR \cite{deRe:talk02034} & di-lepton & 
$\tau_{B^0}=1.557(28)(27)\,\ps$\\
& & $\tau_{B^+}=1.665(26)(27)\,\ps$\\
& & $\tau_{B^+}/\tau_{B^0}=1.064(31)(26)$\\
CDF \cite{CDF:lifetimesEPS03}  
& $B^0\to J\psi(\mu^+\mu^-)K^{*0}$ & $\tau_{B^0}=1.49(6)(2)\,\ps$\\
& $B^+\to J\psi(\mu^+\mu^-)K^+$ & $\tau_{B^+}=1.64(5)(2)\,\ps$\\
 & $B_s\to J\psi(\mu^+\mu^-)\phi$ & $\tau_{B_s}=1.26(20)(20)\,\ps$\\
&  $\Lambda_b\to  J\psi(\mu^+\mu^-)\Lambda$ & $\tau_{\Lambda_b}=1.25(26)(10)\,\ps$\\
D0 \cite{Podesta:March03} & $B^+\to J\psi(\mu^+\mu^-)K^+$ & $\tau_{B^+}=1.76(24)(??)\,\ps$\\
DELPHI \cite{DELPHI:lifetimesEPS03} & charge sec. vtx. & $\tau_{B^0}=1.531(21)(31)\,\ps$\\
& & $\tau_{B^+}=1.624(14)(18)\,\ps$\\
& & $\tau_{B^+}/\tau_{B^0}=1.060(21)(24)$\\
\hline
\hline
\end{tabular}
\end{center}
\caption{\label{tab:newtaub} $b$-hadron lifetime measurements which have appeared since
the CKM Workshop yellow book \cite{Battaglia:2003in}. Many of these results are preliminary.}
\end{table*}
\begin{table}[t]
\begin{center}
\begin{tabular}{ccc}
\hline
\hline
qty & expt & thy\\
\hline
$\tau_{B^0}$ & 1.534(13)\ ps & \\
$\tau_{B^+}$ & 1.652(14)\ ps & \\
$\tau_{B_s}$ & 1.439(53)\ ps & \\
$\tau_{\Lambda_b}$ & 1.210(51)\ ps & \\
$\tau_{B^+}/\tau_{B^0}$ & 1.081(15)& 1.06(2)\\
$\tau_{B_s}/\tau_{B^0}$ & 0.938(35)& 1.00(1)\\
$\tau_{\Lambda_b}/\tau_{B^0}$ & 0.789(34) & 0.90(5)\\
\hline
\hline
\end{tabular}
\end{center}
\caption{\label{tab:sum03taub} World averages of $b$-hadron lifetime measurements, together
with the theoretical predictions reviewed in \cite{Battaglia:2003in}. 
All averages are from \cite{BLG03}, except for  $\tau_{B_s}/\tau_{B^0}$
and $\tau_{\Lambda_b}/\tau_{B^0}$ which are taken to be the ratio of the corresponding
world average lifetimes.}
\end{table}

Taking these new results of
\cite{Aubert:2002ms,Aubert:2002sh,CDF:lifetimesEPS03,DELPHI:lifetimesEPS03}
into account, the B Lifetime Group obtains the averages reported in
\tab{tab:sum03taub} \cite{BLG03}.  
The corresponding lifetime ratios are also given
in
\tab{tab:sum03taub}, together with the theoretical 
predictions reported in \cite{Battaglia:2003in}. 
The experimental
accuracy on $\tau_{B^+}/\tau_{B^0}$, which has reached a stunning
1.4\% thanks to the $B$ factories, is now better than that of the HQE
prediction. And further improvement can be expected from $B$ factories
and the Tevatron soon. The agreement between theory and experiment is
excellent, a clear vindication of the HQE approach.

Agreement for $\tau_{B_s}/\tau_{B^0}$ and
$\tau_{\Lambda_b}/\tau_{B^0}$ is less good, though the discrepancy is
less than two standard deviations. The Tevatron is currently producing
large numbers of $B_s$ mesons and $\Lambda_b$'s such that the error on
these lifetime ratios are expected to be below 1\% by the end of Run
IIa \cite{Rademacker:2003ux}.  This will provide a stringent test of
the HQE and may force theorists to consider penguin contributions,
which are absent in $\tau_{B^+}/\tau_{B^0}$, and which are currently
neglected.

\subsection{Lifetime differences}

On the theory side, the preliminary unquenched, two-flavor results of
the JLQCD collaboration for one of the two matrix elements relevant
for $(\Delta\Gamma/\Gamma)_{B_{d,s}}$ at leading order in $1/m_b$ have
been finalized \cite{Aoki:2003xb}.  These results will not modify the predictions
for $(\Delta\Gamma/\Gamma)_{B_{d,s}}$ reviewed in
\cite{Battaglia:2003in}. Experimentally, the only change comes from the new
measurements of the average $B_{d}$ lifetime, which is used to
obtain $(\Delta\Gamma/\Gamma)_{B_{d,s}}$ from
$\Delta\Gamma_{B_{d,s}}$. The current experimental and theoretical
situation for $(\Delta\Gamma/\Gamma)_{B_{s}}$ is summarized in \tab{tab:dgamma}.

\begin{table}[t]
\begin{center}
\begin{tabular}{ccc}
\hline
\hline
qty & expt (summer 2003) \cite{cite:HFAG} & thy \cite{Battaglia:2003in} \\
\hline
 $(\Delta\Gamma/\Gamma)_{B_{s}}$ & $<  0.29$ at 95\% CL & 0.09(3) \\
\hline
\hline
\end{tabular}
\end{center}
\caption{\label{tab:dgamma} World average   $(\Delta\Gamma/\Gamma)_{B_{s}}$ 
vs. theoretical predictions.}
\end{table}

The status of experimental measurements for
$(\Delta\Gamma/\Gamma)_{B_{s}}$ should change dramatically in the near
future: a statistical uncertainty of $\sim 2\%$ is expected by the
end of Run IIa. It is not clear, however, that theory will be able to
follow. Indeed, the main source of uncertainty comes from $1/m_b$
corrections which are enhanced by a rather large cancellation between
the leading-order contributions.  A calculation of these corrections
requires the non-perturbative estimate of many dimension-7, $\Delta
B=2$ matrix elements which is very challenging. Unfortunately, until
these corrections are calculated with reasonable precision, it is unlikely
that a measurement of $(\Delta\Gamma/\Gamma)_{B_{s}}$ will allow
detection of physics beyond the Standard Model.

\section{Conclusion}
\label{sec:ccl}

We are witnessing very exciting times, with the $B$ factories and the
Tevatron reducing errors tremendously on all of the quantities studied
in our working group. This presents theorist with a great challenge
and will allow for very stringent tests, sending many models to the
grave. The improved experimental accuracy also permits the exploration
of new methods, in which the reliance on non-perturbative calculations
is greatly reduced, such as in the spectral-moments determination of
$|V_{cb}|$.  As this example shows, the close interplay between theory
and experiment is crucial to take advantage of the improved
accuracies. Further gains should be sought by optimizing comparison
between experiment and theory in region of phase space where the
combined errors are minimized, such as in the inclusive and exclusive
determination of $|V_{ub}|$. It is also important to emphasize the
rôle of CLEO-c, which will not only provide accurate branching ratios
necessary for $B$ physics measurements, but will also be very useful
for testing non-perturbative approaches such as lattice QCD and for
calibrating the predictions of these approaches in $B$ physics.

\vspace{0.5cm}
\noindent
{\bf Acknowledgments} 

I wish to thank my co-convenors for their helpful comments on the
manuscript and the participants of Working Group I for many
informative discussions.

\bibliographystyle{my-elsevier}
\bibliography{wg1}

\end{document}